\documentstyle[12pt]{article}
\def\beq{\begin{equation}}
\def\eeq{\end{equation}}
\def\bea{\begin{eqnarray}}
\def\eea{\end{eqnarray}}
\def\bq{\begin{quote}}
\def\eq{\end{quote}}

\def\ga{\left(}
\def\dr{\right)}

\begin{document}
\topmargin -1.0cm
\oddsidemargin +0.2cm
\evensidemargin -1.0cm
\pagestyle{empty}
\begin{flushright}
{CERN-TH.7140/94}\\
PM 93/16 \\
{KEK Preprint 93-184}
\end{flushright}
\vspace*{5mm}
\begin{center}
{{\bf QSSR ESTIMATE OF THE $B_B$ PARAMETER \\
AT NEXT-TO-LEADING
 ORDER}}
 \\
\vspace*{1cm}
{\bf S. Narison} \\
\vspace{0.3cm}
Theoretical Physics Division, CERN\\
CH - 1211 Geneva 23, Switzerland\\
and\\
Laboratoire de Physique Math\'ematique\\
Universit\'e de Montpellier II\\
Place Eug\`ene Bataillon\\
34095  Montpellier Cedex 05, France\\
and\\
{\bf A.A. Pivovarov
\footnote{On leave from
Institute for Nuclear Research of the Russian Academy of
Sciences, Moscow, Russia}}\\
 National Laboratory for High Energy Physics (KEK),\\
1-1 Oho, Tsukuba,
Ibaraki 305,
Japan \\
\vspace*{1.0cm}
{\bf ABSTRACT} \\ \end{center}
\vspace*{2mm}
\noindent
We compute the leading $\alpha_s$ corrections
to the two-point correlator of
the $O_{\Delta B=2}$ operator which controls the $B^0 \bar B^0$ mixing.
Using this result within the QCD spectral sum rules
approach and some phenomenologically reasonable assumptions in the
parametrization of the spectral function,
we conclude
that the vacuum saturation values
$B_B\simeq B_{B^*}\simeq 1$ are satisfied within 15\%.

\noindent

 \vspace*{1.5cm}

\begin{flushleft}
CERN-TH.7140/94 \\
PM 93/16\\
{KEK Preprint 93-184}\\
January  1994
\end{flushleft}
\vfill\eject
\pagestyle{empty}

\setcounter{page}{1}
\pagestyle{plain}

\vskip 2cm

The violation of the CP symmetry is still one of
the most intriguing phenomena in particle physics.
The standard model (SM) offers a pattern for an explanation of
this violation through the complex phase of the unitary
Cabibbo-Kobayashi-Maskawa quark mixing matrix for three generations.
The rigorous experimental test of the pattern
requires knowing the
numerical values of some hadronic matrix elements
at low energies with some non-perturbative QCD methods.

In this paper we discuss an estimate of the $B_d^0$-$\bar B_d^0$
mixing parameter within the QCD spectral sum rules (QSSR)
approach [1,2] along the lines of Ref. [3],
but by including the new $\alpha_s$ corrections and by
taking care on the detailed contributions of different
$B$-like states to the spectral function.
In so doing, we estimate the two-point correlator of the
 $O_{\Delta B=2}=(\bar b_L \gamma_\mu d_L)^2$
operator, where $q_L$ are left-handed quark fields
$$
T(x)
=\langle 0|TO_{\Delta B=2}(x)O_{\Delta B=2}(0)| 0 \rangle .
\eqno (1)
$$
The leading term has a trivial expression in
the configuration space.
Indeed, to leading order in $\alpha_s$, it simply reads
$$
T_0(x)=2 N_c^2 \left(1+{1\over N_c}\right) 16
S'(x,m)S(-x,0)S'(x,m)S(-x,0)
$$
$$
=2 \left(1+{1\over N_c}\right)\mbox{tr}[S(x,m)S(-x,0)]
\mbox{tr}[S(x,m)S(-x,0)]
=2 \left(1+{1\over N_c}\right)\Pi_5(x)\Pi_5(x)
\eqno (2)
$$
where $S(x,m)$ is the free fermion propagator
and $N_c$ stands for the number of quark colours.
The prime means
taking only the part of the propagator
that is proportional to a $\gamma$ matrix.
The function
$\Pi_5(x)=\langle 0|Tj_5(x)j_5(0)| 0 \rangle$
is the two-point correlator
associated to the current $j_5 = \bar b i\gamma_5 d$.

One can also rewrite Eq.~(2) in the form
$$
T_0(x)
=2 \ga 1+{1\over N_c}\dr \Pi_{\mu\nu}(x)\Pi^{\mu\nu}(x),
\eqno (3)
$$
where
$\Pi^{\mu\nu}(x)=\langle 0|Tj_L^\mu(x)j_L^\nu(0)| 0 \rangle$
and
$j_L^\mu=\bar b_L \gamma^\mu d_L$,
which has the following
Lorentz decomposition in $x$-space
$$
\Pi^{\mu\nu}(x)
=(-\partial^\mu\partial^\nu+g^{\mu\nu}\partial^2)\Pi_T(x^2)
-\partial^\mu \partial^\nu\Pi_L(x^2).
$$
Formulae (2) and (3) demonstrate an explicit factorization
of the correlator (1) in the configuration
space to leading order in $\alpha_s$.

The dispersion representation in $x$-space for any
two-point correlator $\Pi_j(x)$ ($j=T,L,5$) has the form
$$
i\Pi_j(x^2)=\int_{s_j}^\infty r_j(s)D(x,s)ds,
\eqno (4)
$$
where $D(x,s)$ is a free boson
propagator with the ``mass" $\sqrt s$. The spectral functions $r_j$,
read to leading order in $\alpha_s$:
$$
r_L^{(0)}(s)=N_c{1\over 16 \pi^2}z(1-z)^2,
{}~~r_T^{(0)}(s)=N_c{1\over 16 \pi^2}{1\over 3}(1-z)^2(2+z),
$$
$$
r_5^{(0)}(s)=m_b^2 N_c {1\over 16\pi^2} 2 {(1-z)^2\over z},
\eqno (5)
$$
where $z=m_b^2/s$, $m_b$ is the $b$-quark pole mass.

One can define the spectral function $\rho(s)$ of the
full correlator $T(x)$ in the same way as in Eq.~(4) and express
it, to first order in $\alpha_s$,
in terms of the spectral functions $r_j(s)$ associated
to the two-line correlators. Therefore
$$
\rho(s)=\int r_1(s_1)r_2(s_2)\Phi(s;s_1,s_2)ds_1ds_2,
\eqno (6)
$$
where
$$
\Phi(s;s_1,s_2)={1\over 16\pi^2 s}
\sqrt{(s-s_1-s_2)^2-4 s_1 s_2}
\eqno (7)
$$
is the two-body phase-space factor, and the concrete form
of the spectral functions $r_j(s)$ entering Eq.~(6)
depends on the representation chosen for the correlator
$T(x)$ (as in Eqs.~(2,3) for $T_0(x)$).
The integration region in Eq.~(6) is determined by the properties
of the phase-space factor (7) for corresponding
representation of the whole correlator;
$\Phi(s;s_1,s_2)$ is supposed to be equal to zero within
kinematically forbidden regions.

To leading order in $1/N_c$, one can write the correlator $T(x)$
as a product of two two-line correlators
given in Eq.~(4). It is worthwhile to notice that
this decomposition is gauge-invariant
and finite, i.e. it does not require any renormalization
that is a reflection of the vanishing of
the operator $O_{\Delta B=2}$
anomalous dimension in leading order in $1/N_c$.

Including $\alpha_s$ corrections, the spectral function
in the factorization approximation reads:
$$
\rho_{fact}(s)=\rho_0(s)
(1+\Delta \rho_f(s)),
\eqno (8)
$$
where $\rho_0(s)$ generates the leading-order term
and $\Delta \rho_f(s)$ is a (properly normalized)
factorized correction in the $\alpha_s$ order.
The colour structure is the following
$$
\rho_0(s)=N_c^2\left(1+{1\over N_c}\right)\tilde\rho_0(s),
{}~~~\Delta \rho_f(s)=C_F{\alpha_s\over \pi}\Delta
\tilde \rho_f(s),
$$
where $C_F=(N_c^2-1)/2N_c$ for the $SU(N_c)$ colour
group and the quantities with the tilde contain
no explicit $N_c$ dependence.

The representation of the spectral density $\rho_{fact}(s)$
through those of two-line correlators for
the vector-like decomposition in Eq.~(3) is:
$$
\rho_{fact}(s)= \int ds_1ds_2
\Phi(s;s_1,s_2)
\{ \ga {s^2_{12}\over 4}+2s_1 s_2\dr r_T(s_1)r_T(s_2)
$$
$$
+\ga{s^2_{12}\over 4}-s_1 s_2\dr
\ga r_T(s_1)r_L(s_2)+r_L(s_1)r_T(s_2)\dr
+{s^2_{12}\over 4}r_L(s_1)r_L(s_2 \}
$$
where $s_{12}=s-s_1-s_2$.
The non-factorizable corrections are of the $1/N_c$ order
and the full spectral density can be written
in the form
$$
\rho(s)=
\rho_0(s)\left(1+\Delta \rho_f(s)+\Delta
\rho_{nf}(s)\right)
\eqno (9a)
$$
and
$$
\Delta
\rho_{nf}(s)=
{C_F\over N_c+1}{\alpha_s\over \pi}\Delta \tilde \rho_{nf}(s)
=
{1\over 2}\left(1-{1\over N_c}\right){\alpha_s\over \pi}
\Delta \tilde\rho_{nf}(s).
\eqno (9b)
$$
One of the diagrams contributing to the non-factorizable
part of the whole spectral density is given in Fig.~1.

Several comments are in order here. The decomposition (9)
is gauge-invariant. The same is true for
the anomalous dimension of the operator $O_{\Delta B=2}$.
We use four-dimensional algebra of Dirac's $\gamma$ matrices
throughout the computation.
This seems quite natural because it allows one
to make Fierz rearrangements freely, which
is crucial for establishing the validity of
factorization.  At the same time this approach simply implies
the special choice of the renormalization scheme.
So, this scheme is not the standard
$\overline{\rm{MS}}$ one and
it is not the real dimensional reduction as used in [4]
either.
To the considered order in $\alpha_s$,
however, the scheme dependence reduces to
a certain choice of the normalization parameter $\mu$.
In principle, one can fix the scheme by direct comparison
of our results
with the massless limit for corresponding correlator
[5] or, to put it another way, by
comparing the non-logarithmic parts of corrections in the massless
limit one can obtain the relation between our parameter $\mu$
and
the corresponding $\mu_{\overline{MS}}$ or $\mu_{DimRed}$.
Thus, the answer in any desired scheme can be easily recovered
by considering the massless limit.
In the present
paper
we
do not dwell on this point.

We have computed numerically the spectral density $\rho(s)$
in the first non-leading
order in $\alpha_s$. We analyse our results,
paying special attention to the presentation of the
entire spectral density
$\rho(s)$
as a sum of factorizable and
non-factorizable pieces, as in Eq.~(9).
This decomposition is useful from the theoretical point of
view. It is also interesting from the $1/N_c$ analysis,
and
it is quite convenient technically.  As for the factorizable
part of the spectral density (Eq.~8) the analytical expressions for
two-line spectral functions $r_{T,L}(s)$
are well known
to  first order in $\alpha_s$,
and we
use them as they have been
given in Refs. [6,7].
For the non-factorizable part of the spectral density, we have to
compute the gluon-exchange diagram both for unequal mass
lines and for equal mass lines of a two-line correlator.  We
have done it using the REDUCE system of analytical
computation [8]
and the table of two-loop integrals given in
[6,7].
After using a Fierz transformation, these non-factorizable
diagrams
can be represented in form analogous to that of the factorizable
diagrams, i.e. the corresponding
analytical expression is given by
a product of two traces in Dirac's
indices.
The transformed diagrams are given in Fig.~2.

For diagrams with a gluon connecting $b$ or $d$-quarks in the loop
(Fig.~2(a,b)), the resulting representation can be rendered
into a product of two scalar spectral densities of two-line correlators.
The relevant part of the spectral density reads:
$$
r_m(s)={1\over 16\pi^2}s\left\{(2(1-2 z)v
+C_F{\alpha_s\over \pi}
\left(-4 v (1-2 z)({1\over \epsilon}+2\ln{\mu^2\over m_b^2}) \right.\right.
$$
$$
+4(1-2 z)(1-v)\ln z+(1+v)(4zv+(3+v)(z-1))\ln{1-v\over 1+v}
$$
$$
\left.\left.+(1-2z)^2\phi(u)+v(18z-13)+16z-8\right)\right\},
$$
where $\epsilon=2-D/2$, $D$
is the dimensional regularization parameter,
$v=\sqrt{1-{4 m_b^2\over s}}\equiv\sqrt{1-4z}$,
$\phi(u)=8(Sp(u^2)-Sp(-u))+4(2\ln(1-u^2)-\ln(1+u))\ln u $,
$u=(1-v)/(1+v)$,
$$
Sp(u)=-\int_0^u{\ln(1-t)\over t}dt.
$$
For a zero mass correlator
($d$-quark), the spectral density has the form
$$
r_0^(s)={1\over 16\pi^2}s\left\{2+C_F{\alpha_s\over \pi}
\left(-4 ({1\over \epsilon}+2\ln {\mu^2\over s} )-21\right)\right\} .
$$

The corresponding representation
for diagrams, with different quark flavours tied with a gluon in the
loop,
is of the vector type. The
nonfactorizable first order corrections to the spectral densities
are:
$$
r_{L,T}(s)=r_{L,T}^{(0)}(s)+C_F{\alpha_s\over \pi}r_{L,T}^{(1)}(s),
$$
$$
r_L^{(1)}(s)=z(1-z)^2({1\over \epsilon}+2\ln{\mu^2\over m_b^2})
-2z(1-z)^2\ln(1-z)\ln({z\over 1-z})
$$
$$
+z(1-z)^2(4z^2-3z+{1\over 2})\ln({z\over 1-z})
+4z(1-z)^2 Sp(-{z\over 1-z})
+z(1-z)(2z-3)\ln(1-z)
$$
$$
z(4z^2-7z+{3\over 2})\ln z
+{65\over 12}z^3-{17\over 2}z^2+{21\over 4}z-{2\over 3},
$$
$$
3 r_T^{(1)}(s)=(z+2)(1-z)^2({1\over \epsilon}+2\ln{\mu^2\over m_b^2})
-2(z+2)(1-z)^2\ln(1-z)\ln({z\over 1-z})
$$
$$
+(1-z)^2(2z^3-2z^2+{5\over 2}z+2)\ln({z\over 1-z})
+4(z+2)(1-z)^2 Sp(-{z\over 1-z})
$$
$$
+z(1-z)(2z+1)\ln(1-z)
+(3z^3+2z^2-{9\over 2}z+2)\ln z
+{41\over 12}z^3+{3\over 2}z^2-{75\over 4}z+{46\over 3}.
$$

The whole computation can be subjected to a
powerful test consisting in the cancellation of
divergences by the one-loop renormalization constant of the
operator $O_{\Delta B=2}$, which is known
to be $Z_{O_{\Delta B=2}}=1-3(1-1/N_c){\alpha_s\over 4\pi\epsilon}$.
We checked this cancellation explicitly.

The QCD results for the spectral density are given in Table 1.

At larger $s$ (say, $s\ge 5 m_b^2$),
where the perturbative QCD result makes sense,
one can notice that the
non-factorizable correction dies out faster than the
factorizable one, so that the full correction comes mainly
from that of the two-line correlator.
Therefore, at sufficiently large $s$, the factorization
(in the sense of Eqs.~(2) and (3)) becomes exact
and the `formal' $1/N_c$ suppression
of non-factorizable corrections becomes numerically valid.

Using the QSSR approach
for estimating $B_B$, we can work with the moments:
$$
M_i(s_{th})=\int_{4m^2_b}^{s_{th}}\rho(s)s^{-i}ds,
\eqno (10)
$$
which can be decomposed as:
$$
M_i(s_{th})=M_i^0(s_{th})
\left(
1+\Delta M_i^f(s_{th})+
\Delta M_i^{nf}(s_{th})
\right),
$$
according to the corresponding decomposition of
the spectral density, Eq.~(9a).

We show in Table 2 the strength
of the non-factorizable correction for $s\ge 5 m_b^2$
for the following set of input parameters:
$\Lambda^{(5)}_{\overline{MS}}=175~$MeV,
$m_b=4.6~$GeV,
and with the one-loop expression for the strong coupling constant
$\alpha_s(\mu)=6\pi/23\ln(\mu/\Lambda^{(5)}_{\overline{MS}})$
at the point $\mu=m_b$.
For given regions of integration
in Eq.~(10), the moments
of the factorizable spectral density are
practically independent of the power $i$
of the weight function $s^{-i}$.
They change by less than 5\% for $i$=1-10.
The non-factorizable correction changes its sign within
the integration region for $s\ge 5.5 m_b^2$,
and its moments are more sensitive to the power $i$.
One can notice that the non-factorizable
correction does not exceed a 15\% level with respect to the full
factorized spectral density at sufficiently large $s$.
The full order-$\alpha_s$ corrections (factorizable+non-factorizable)
can be quite large for both the spectral density itself and its moments
depending on the energy $s$ as they
can reach a magnitude of 100\% with respect to the leading
term. Hopefully, the non-factorizable corrections,
measured in terms of the fully factorized
(lowest order + $\alpha_s$ terms) spectral density
are, nevertheless, still moderate. However,
all the factorizable corrections
to the correlator (1),
no matter how large they are, can be absorbed into the calculation
of the decay constant $f_B$ from the two-point correlator with two
quark lines, in such a way that the relevant corrections to the $B_B$
parameter are only due to the non-factorizable ones.

Contrary to the QCD part, the estimate of the phenomenological
contributions to the correlator is much more involved.
For the analysis to be performed one might make the following
phenomenological assumptions:

\noindent
-- the minimal choice of operator:
$O_{\Delta B=2}=g_{\Delta B=2}\partial_\mu B \partial^\mu B$,
can give a good description of the spectral function.

\noindent
-- the contributions of
the $BB$ and
$B^* B^*$ pairs to the spectral function are equal
due to their approximate degeneracy
and to the equal values of their decay constants [9]
(this feature is fully satisfied in the large $m_b$ limit);

\noindent
--we do not have  exotic contributions due to
a singlet mixture of colored states or due
to some four-quark-like states which may
restore the $N_c$-structure of the correlator
in order to match the proper $(1+1/N_c)$
construction entering the standard definition
of the $B_B$ parameter.

Within these phenomenological assumptions,
we can conclude that the effect of inclusion of the perturbative
corrections into the QCD part of the correlator on the $B_B$ parameter
is reasonably small and the change is not larger than $15\%$.
The absolute value
of $B_B$ however cannot be unambiguously established
using the perturbative part of the correlator only.
Assuming that non-perturbative non-factorizable corrections are small
and the factorizable corrections are properly taken into account
through $f_B$ we find after correcting for exotic contributions
in theoretical part of the correlator
$$
B_B \simeq B_{B^*}
\simeq (1.00\pm 0.15)
$$
where we have not explicitly written the small effect due to the
anomalous dimension of the operator $O_{\Delta B=2}$.

An improvement of our result needs an explicit quantitative check of
the previous phenomenological assumptions.
Most probably, the QSSR approach based on the three-point function
[10] may be more appropriate, as it does not need some of the assumptions
which we have used for the two-point function. Unfortunately,
one has to face, in this case, a highly involved computation of the
non-recursive three-loop non-factorizable diagrams
in order to obtain the $\alpha_s$ corrections.

{\bf Acknowledgement}

S.N. would like to thank A. Pich for discussions.
A.A.P. acknowledges the kind hospitality extended to him at
the Laboratoire de Physique Math\'ematique of Montpellier,
where the main part of the present
computation
has been done
and thanks the Centre International des Etudiants et Stagiaires (CIES)
of the European Community for financial support.
The work of A.A.P. is partly supported by Japan Society for the Promotion
of Science (JSPS).

\newpage
\begin{table}
\begin{center}
\begin{tabular}{|c|c|c|c|} \hline

$s/m_b^2$   &$\Delta\rho_f$ &$\Delta\rho_{nf}$
&$\Delta\rho_{nf}/(1+\Delta\rho_f)$\\ \hline

5.5 &1.03   &0.02     &0.01 \\ \hline
6.0 &0.95   &0.21     &0.11  \\ \hline
6.5 &0.89   &0.29     &0.15  \\ \hline
7.0 &0.84   &0.32     &0.17  \\ \hline
\end{tabular}
\caption{Normalized spectral densities}
\vspace{3cm}
\begin{tabular}{|c|c|c|c|c|} \hline

$i$&$s_{th}/m_b^2$&$\Delta M_f$ &$\Delta M_{nf}$ &$\Delta
M_{nf}/(1+\Delta M_f)$\\ \hline

0&5.5   &1.07 &-0.16 &-0.08  \\ \cline{2-5}
 &6.0   &0.99  &0.11  &0.06  \\ \cline{2-5}
 &6.5   &0.93  &0.23  &0.12  \\ \cline{2-5}
 &7.0   &0.88  &0.29  &0.15  \\ \hline

5&5.5   &1.08 &-0.21 &-0.10  \\ \cline{2-5}
 &6.0   &1.00  &0.08  &0.04  \\ \cline{2-5}
 &6.5   &0.94  &0.20  &0.11  \\ \cline{2-5}
 &7.0   &0.89  &0.27  &0.14  \\ \hline

10&5.5   &1.09 &-0.27 &-0.13  \\ \cline{2-5}
  &6.0   &1.01  &0.02  &0.01  \\ \cline{2-5}
  &6.5   &0.96  &0.16  &0.08  \\ \cline{2-5}
  &7.0   &0.91  &0.23  &0.12  \\ \hline
\end{tabular}
\end{center}
\caption{Normalized moments of spectral densities}
\end{table}

\end{document}